# Switchable Crystalline Islands in Super Lubricant Arrays

Youngki Yeo[1], Yoav Sharaby[1], Nirmal Roy[1], Noam Raab[1], Watanabe Kenji[2], Takashi Taniguchi[3], Moshe Ben Shalom[1]

[1]School of Physics and Astronomy, Tel Aviv University, Tel Aviv, Israel
[2]Research Center for Electronic and Optical Materials, National Institute for Materials Science, Tsukuba, Japan
[3]Research Center for Materials Nanoarchitectonics, National Institute for Materials Science, Tsukuba, Japan

Expanding the performance of field effect devices is a key challenge of the ever-growing chip industry at the core of current technologies[1,2]. A highly desired nonvolatile response in tiny multiferroic transistors is expected by electric field control of atomic movements rather than the typical electronic redistribution[3]. Recently, such field effect control of structural transitions was established in commensurate stacking configurations of honeycomb van der Waals (vdW) polytypes by sliding narrow boundary dislocations between oppositely polarized domains[4–7]. The interfacial ferroelectric response, however, relied on preexisting boundary strips between relatively large micron-scale domains, severely limiting practical implementations[4,8,9].

Here, we report the robust switching of single-domain polytypes in nm-scale islands embedded in super lubricant vdW arrays. We etch cavities into a thin layered spacer and then encapsulate it with parallel functional flakes. The incommensurate flakes above/under the spacer sag and touch at each cavity to form uniform crystalline islands free from interlayer deformations. By imaging the polytypes' ferroelectric response, we observe reversible nucleation and annihilation of boundary strips and geometry-adaptable hysteresis loops. Using mechanical stress, we accurately position the boundary strip, modify the interlayer twist angle, and nucleate intermediate polar domain patterns. By precisely designing the size, shape, symmetry, and distribution of the islands in these Super Lubricant Arrays of Polytype (SLAP), we envision numerous device functionalities and "slideTronics" applications[10]. These range from ultra-sensitive detectors of atomic-scale shifts to nonvolatile multi-ferroic tunneling transistors with tunable coercive switching fields, and even elastically-coupled memory cells for neuromorphic architectures.

**Main:** Unlike electronic transitions, crystalline structural transitions are challenging to control due to considerable energy barriers associated with breaking solid bonds at ambient conditions, away from the melting temperature. Nevertheless, some materials may exhibit practical transitions between amorphous and crystalline orders in response to external stimuli like optical or electric pulses[11]. Switching the discrete periodic symmetries in these "phase-change" materials directly impacts their collective lattice excitations and numerous subsequent properties. Thus, electric control of structural transitions enables, in principle, rapid switching of intrinsic responses such as light emission, conductivity, and magnetic order in so-called multiferroic devices[12,13].

Recent experiments demonstrated exceptional electric field switching between vdW polytypes with discrete commensurate stacking configurations that break inversion and mirror symmetry[14–17]. Owing to relatively weak interlayer adhesion and high planar stiffnesses, partial dislocations strips between domains with opposite structural and polar orientations elongate and slide[4] to expand better-stable configurations of co-aligned polarization $P_z$. The stacking-fault dislocation strips in polytypes of honeycomb graphene[18], hexagonal boron nitride[19] (h-BN), or transition metal dichalcogenides[20] (TMDs), are ~ 30 atoms wide as determined by the Burgers vector in the partial dislocation (one bond length), the potential well energy in the commensurate state (~ one meV per atom), and the planar elastic modulus[21] (~ 1 TPa). The stacking misalignment, extra interlayer separation, and planar strain energy in the boundary set an energy barrier of ~1 eV per nm strip length[19]. Thus, nucleating the boundary strips, an essential step for structural transition, is restricted at room temperature (below the turbostratic transition[22]), even if cutting the structure into small nanoscale islands and applying local external stimuli[4]. Here, the open dangling bonds at the physical edges of the layers tend to zip the layers together and restrict interlayer motions[8]. As a result, previous electric hysteresis measurements have relied on preexisting dislocation strips in micrometer-size structures[4,8,9,23,24]. Additional challenges arising in mechanically assembled interfaces are uncontrollable twists and stiff moiré networks that tend to freeze the sliding and restrict local switching. Moreover, after removing the external field, the rigid strip network pulls the moiré pattern back to its initial dimensions and eliminates hysteretic memory response[8,25].

The Super Lubricant Arrays of Polytypes (SLAP) reported here overcome these challenges by embedding tiny commensurate islands in an incommensurate super lubricant medium, enabling purely electrical domain nucleation and subsequent switching of single-polytype islands in numerous nanoscale devices. Given the various electronic phases of vdW polytypes, such as cumulative polarizations[26–28], superconductivity[29], spin[30], orbital[31], and topological orders[32], the established structural switching of SLAP islands by electric field offers remarkable multiferroic control mechanisms for subsequent electronic phases[10]. Beyond highly desired islands that snap efficiently into single structural domains, controlling single boundary strips [33–36], interlayer twist angles[37–42], and topological domain patterns[43–45] in each island enables further device functionalities. Here, overcoming the dislocation energy cost, and the corresponding anhelation instabilities requires considerable pinning barriers that compromise the efficient electric switching. We address these challenges by exploring the mechanical response of several SLAP devices with different active layers and geometries.

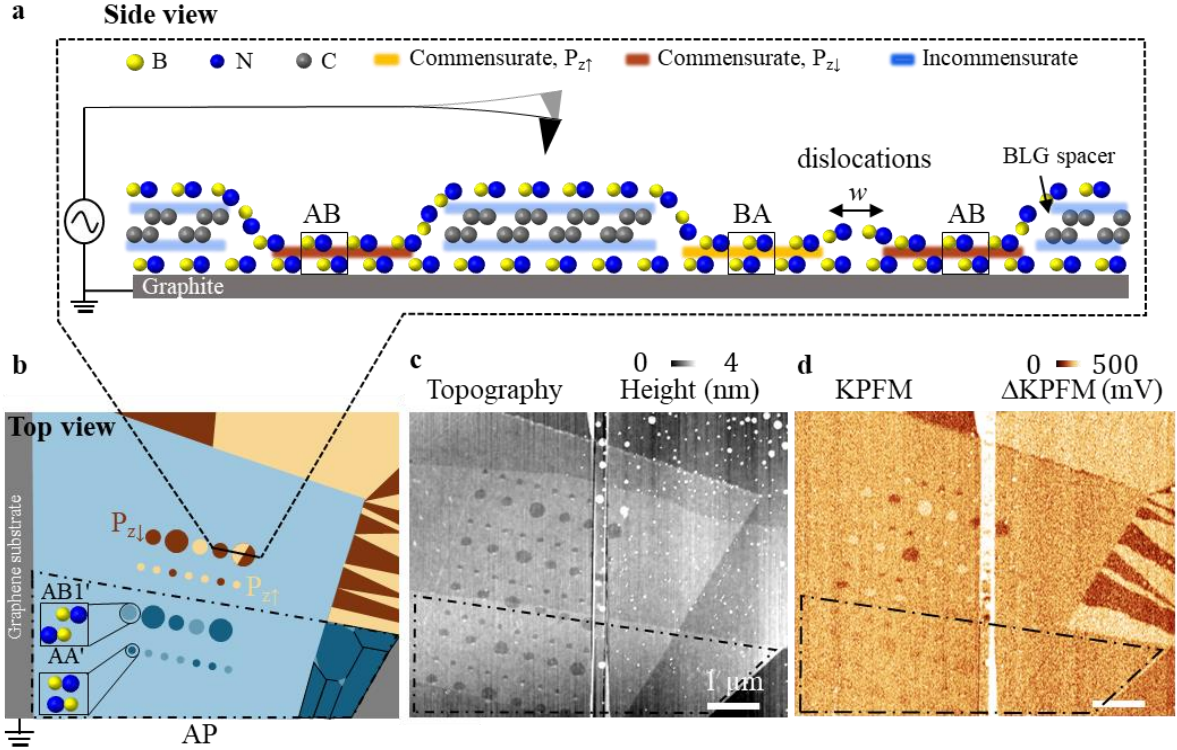

**Fig. 1: Super-Lubricant Array of vdW Polytypes (SLAP). a,b**, Sideview, and Top view device illustrations. The line cut shows a pair of cavities etched in a bilayer graphene spacer (identical carbon spheres) and encapsulated by two parallel *h*-BN layers (yellow and blue spheres). Commensurate *h*-BN interfaces of AB/BA polytype configurations at the cavities are shaded with dark/bright brown color for down/up internal polarizations. Lubricant incommensurate interfaces with the spacer are cyan-shaded. The right-side cavity island shows mixed BA/AB domains separated by a *w*-wide partial dislocation strip. The dashed-dotted line borders a region of antiparallel AA' and AB1' polytypes[46] islands marked by dark blue/bright blue colors, respectively. **c**, Topographic image of Device 1 with two nm thick parallel *h*-BN flakes. Note the brighter (one layer thicker) surface below the dashed-dotted line. **d**, Corresponding electric surface potential map by Kelvin probe force microscopy (KPFM).

**Device concept - sliding polytypes in lubricant cavities:** The main idea behind our SLAP devices is to 1) Squeeze the active commensurate interface into small cavities without physically cutting the functional layers and introducing open edge bonds. 2) Use lattice mismatched or considerably twisted layered spacers to support super-lubricant dynamics between the active cavities.

Since interfaces of lattices mismatch/twisted layers by more than 2% / a few degrees are not commensurate, the "not-active" inter-cavity regions can facilitate exceptionally high sliding lubricity with record low ~$10^{-5}$ friction coefficients[47–49]. Via this super lubricant medium, the stress field can spread to reduce the elastic deformation and mediate long-range elastic interactions, which are typically restricted by any pinning barrier. For example, the pining potential wells in the commensurate stacking state (that prevents super lubricity) confine the interlayer shifts within the boundary dislocations strip only. Here, the elastic energy cost ~

$\frac{1}{4}k\frac{a^2}{w}$ ($k\sim 150$ N/m the shear stiffness, $a\sim 0.14$ nm the bond length) is competing with the misalignment energy $\sim wE_{sp}$ (with $E_{sp}$ the pinning energy per unit area $\sim 1$meV/atom) to set a strip width $w = \frac{a}{2}\sqrt{\frac{k}{E_{sp}}} \sim 7$ nm in graphene[18] and $h$-BN strips[50]. However, on the super lubricant interfaces with orders of magnitude lower friction and pining potentials, the strain relaxation length can substantially exceed $w$ to lower the elastic deformation energy and spread the inter-cavity interactions beyond the present few nm-scale interactions between adjacent dislocation strips[10,51]. Notably, the dislocation networks and the topological constraints associated with local atomic relaxations and global interlayer twists are eliminated outside the cavities[10,18]. Rather, boundary dislocations may nucleate/annihilate independently in each isolated island with an energy budget determined only by the cavity's dimensions and the remaining pinning potentials. Altogether, the SLAP concept provides freedom to design the shape and geometry of each cavity down to the nm scale while simultaneously controlling their mutual elastic and electronic interactions by designing the symmetry and spread of the array.

Figure 1 illustrates cross-sectional and top views of the conceptual array using a graphene spacer and functional $h$-BN flakes. We etch circular cavities into a graphene bilayer (BLG, see the identical carbon spheres in Fig.1a) using the electrode-free local anodic oxidation (LAO) method[52] (see methods). Then, we encapsulate the BLG with $h$-BN layers (blue and yellow spheres, respectively) that sag to touch at the cavity position, forming a parallel and commensurate AB or BA interfaces (dark/bright brown shaded). These stable polytype configurations break inversion and mirror symmetry (see rectangular frames with a line-cut illustration of a unit cell along the armchair direction), inducing vertical polarization at this active interface. Outside the cavity, on the other hand, the $h$-BN/graphene interfaces are incommensurate owing to lattice mismatch and finite twist angle (blue-shaded interfaces). Fig.1b shows a top view of a cavity array design with various cavity diameters and inter-cavity spaces. Outside the blue-shaded spacer (top right side), triangular domains of AB and BA configuration form, with opposite structural orientations and internal polarizations $P_z$ (noted by bright/dark colors) as in our previously reported devices[4]. Islands of non-polar AA' and AB1' polytypes that preserve inversion symmetry (see framed illustrations) appear in regions of antiparallel functional layers as expected for an extra (rotated) interfacial layer[16] (framed in a dashed-dotted black line).

**Detecting polytypes by imaging interfacial ferroelectricity:** To monitor the cavity polytypes, as a ferroelectric-case study, we use flakes of $h$-BN or $WSe_2$ as the functional encapsulating crystals (see Device 1-3 details in Table S1, Fig. S1 and SI. 1). A parallel commensurate stacking of these binary compounds spreads the electrons unequally between the top and bottom layers, inducing interfacial-confined polarization $P_z$ which is resilient to nearby interfaces configurations[4]. Conversely, $P_z$ nullifies for interfacial twists beyond a few degrees that decouple the functional layers. Notably, each interfacial shift by a single inter-atomic distance switches the untwisted AB stacking to BA and vice versa, which is equivalent to switching the interface and $P_z$ upside down[14] (see rectangular frames in Fig. 1a). Hence, monitoring $P_z$ is an exceptional sensor for minute interfacial shear motions. Opposite $P_z$ orientation in adjacent AB/BA domains modifies the electric surface potential by $\Delta V \sim 240, 120$ mV for *h*-BN and WSe$_2$, respectively[4,6,16], allowing us to distinguish each polytype configuration deterministically. Figure 1c presents an atomic force microscope (AFM) topography map of the device structure illustrated in Fig.1a,b. We cut straight borders, designed 300, 200, 100, and 20 nm cavity diameters in the graphene spacer, and used 2 nm thick functional *h*-BN flakes (see darker circles, and further details in SI. 2 and Fig. S2).

The surface potential map, measured by Kelvin Probe Force Microscopy (KPFM, see methods), is presented in Fig. 1d. Outside the spacer, at the top and right sides of the map, bright/dark domains differing by 240 mV potential steps confirm the formation of polar AB and BA interfaces. Notably, the same bright/dark circles at the cavity's location reveal uniform single-domain polytypes embedded in uniform spacer potential. A similar response is observed in Device 2 with a monolayer graphene spacer (Fig. 2, Fig. S1,3-7) and Device 3 with a trilayer graphene spacer and WSe$_2$ monolayers as the functional flakes (Fig. 3, Fig. S1). We note that all cavities above the dashed dark line in Fig.1d appear either bright or dark. Conversely, below this line, the potential at the cavities and outside the spacer is measured to be the average potential as expected for non-polar antiparallel interfaces. The topography map in this section confirms an additional *h*-BN layer in the bottom flake (7/6, 7/5 layers in the top/bottom flake below/above the dashed line, respectively) and the formation of mirror symmetric AA' or AB1' interfaces[4,16] (as in the naturally grown 2H flakes, see framed illustration in Fig. 1b). Altogether, the topography and potential maps confirm commensurate interface formations in pristine cavity arrays spanning many µm$^2$ regions, released from external contaminations, even at the circular edges of the spacer.

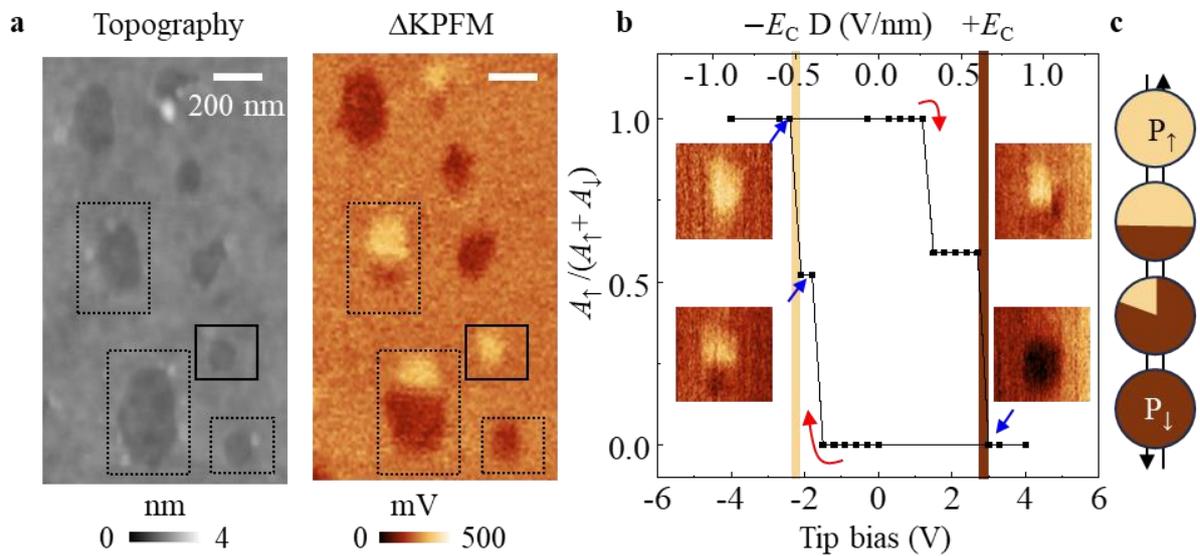

**Fig 2: Electric field-induced dislocation nucleation and annihilation. a**, AFM topography and surface potential maps of Device 2 before electric field poling. **b**, The relative bright domain area ($A_↑$) in the 150 nm cavity marked by a solid line frame in **a,** as a function of the tip bias during the electric poling scan before imaging. The coercive switching fields are indicated by bright/dark bars. Full hysteresis loop maps of all marked cavities are shown in Fig. S3,4. **c**, Illustration of intermediate domain structures observed in the experiments (see Fig. S5).

**Boundary strip nucleation and annihilation for ferroelectric Switching:** To explore the cavities switching dynamics, we applied external displacement fields between the AFM tip and a bottom graphite electrode using bias contact scanning mode (methods). A set of poling and then imaging maps are taken above representative cavities with different diameters (Fig. 2a). Figure 2b shows surface potential maps of a 150 nm wide cavity after increasing/decreasing the poling bias by 0.3 V steps, corresponding to electric displacement field steps of 0.05 V/nm. Notably, a new dislocation strip nucleates at ~ ±0.3 V/nm and then slides at a coercive field $E_c$ ~ ±0.6 V/nm to entirely switch the cavity between uniform single-domains of up/down polarization. While previously reporting smaller coercive fields of ~ 0.3 V/nm, we note that these former experiments[4] could only slide preexisting dislocation strips without nucleating new strips up to breakdown displacement fields as high as 1.5 V/nm.

To analyze the hysteresis loop and the intermediate polarization states, we plot the relative coverage of up (bright) domains, $A_\uparrow/(A_\uparrow + A_\downarrow)$. A complete set of maps is shown in SI. 3 and Fig. S3 and was repeated over five switching cycles. The hysteresis loops of three additional cavities marked in dashed frames in Fig. 2a are presented in SI. 3 and Fig. S4. While the 150 and 250 nm cavities show similar hysteresis windows, we could only switch the 350 nm cavity partially under electric displacements as high as 1 V/nm (before the sample is damaged). In all cavities, the center of the hysteresis loop shifts to positive or negative potentials with no apparent sign preference. A preferred bias shift is expected for vertical electric fields from the asymmetric tip/graphite electrodes or flexoelectric fields (due to pressure gradients from the tip curvature)[53]. Therefore, we attribute the random shifts to elastic coupling with adjacent cavities and remote pinning by nearby contaminants. We also find strip elongations aligning with the smaller width axes of the oval-shaped cavities (see the two bottom left cavities in Fig.2a). Such tendency to shrink the strip length and to snap many cavities into single domains (see Fig.1d) testifies that releasing the strain to the lubricant interfaces is energetically favorable. Occasionally, we find intermediate triangular domains snapping into 120- or 180-degree head angles rather than straight boundary strips (see the Illustration in Fig. 2c, SI. 4, and Fig. S5). Such a configurable intermediate domain pattern appearing at confined ferroelectric islands is raising interest for topological texture applications and was recently reported in oxide nanoplates[45].

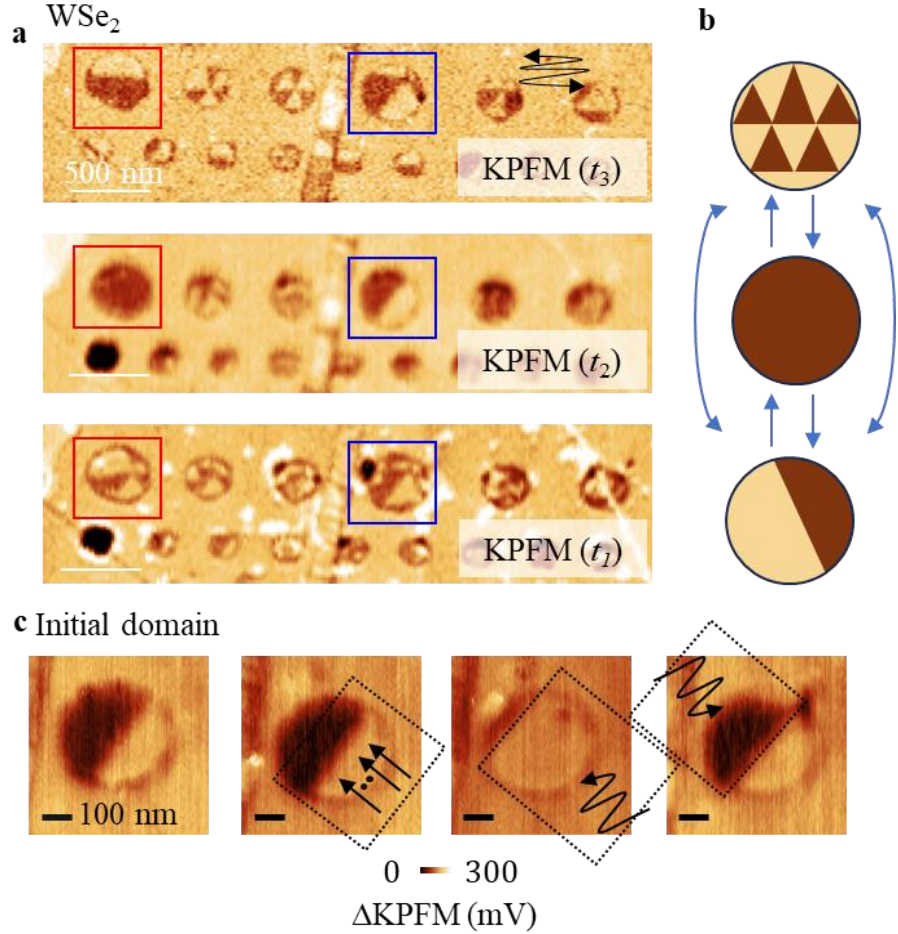

**Fig 3: Mechanical nucleation and reconfiguration of confined domain patterns. a**, Surface potential maps of Device 3 with functional $WSe_2$ monolayers before and after contact mode scans at a tip force of 300 nN. Black arrows show the orientation of the slow and fast contact scanning axes before the non-contact imaging. **b**, Schematic illustration of the cavity topologies observed, including single, two, and multi-domains of twisted moiré patterns. **c**, Surface potential maps taken after pressure scanning the regions marked in dashed lines. Scanning the fast axes perpendicular to the boundary strip in the forward direction only (straight arrows) does not change the final pattern, while back and forward scanning of the fast axes along the strip (armchair) direction (curved arrows) controls the boundary position accurately and reversibly.

**Mechanical switching and interlayer twists:** External stress fields may also nucleate and reconfigure domain patterns in confined ferroelectric structures. 3D ferroelectrics are known to switch under local stress gradients owing to a flexoelectric response[54]. Beyond the flexoelectric stimuli, switching of layered polytypes is expected even under planar stress fields orthogonal to the internal polarization[55]. For example, the scanning AFM tip is known to manipulate preexisting dislocation strips between non-polar graphitic polytypes for vertical loads as low as 400 nN (refs.[33,36]). However, dislocation nucleation and domain switching remain limited and required intensive mechanical loads exceeding 40 μN (ref.[35]). Nucleating dislocations in twisted interfaces is an even greater challenge due to the rigid moiré network and topologically defined number of alternating domains for a given global twist angle[18]. Once the planar relaxation clicks the layers into commensurate domains, the external stress impact becomes irreversible[39,40].

To assess the mechanical switching dynamics in SLAP devices, we scanned the arrays in AFM contact mode while applying vertical pressures of 50 to 300 nN and keeping the slow axis scan direction parallel to the underlining armchair direction (see methods). Figure 3a presents three surface potential maps of Device 3 (with $WSe_2$ as active layers) measured before and after applying a vertical pressure of 300 nN at times – $t_1$, $t_2$, $t_3$. Notably, the domain patterns modify between the scans in nearly all cavities. The red/blue rectangles mark relatively large cavities in which triangular domains are tuned to a single/double domain pattern and back, respectively, as illustrated in Fig.3b. We find a threshold load for domain nucleation and motion of ~ 200 nN (Fig. 3), corresponding to a pressure of ~161MPa (SI. 5 and Fig. S6) on a ~1000 $nm^2$ effective tip diameter, and a planar shear force of ~30nN measured by the horizontal deflections of the tip (see Fig. S8).

We note the polydomain structures observed with several triangular domains, rather than the uniform islands observed in Devices 1,2 (Fig. 1,2). Typically, the triangles dimensions indicate the twist angle between the active layers (100 nm domains correspond to ~0.1°) before the stacking and the planar atomic relaxations happens[10]. In the SLAP devices however, the tendency to observe unform domains suggests a counter-twist response in high-quality devices with barrier free superlubricant medium (which might be compromised by the $WSe_2$ quality compared to BN). Alternatively, for sufficient twist angles the polydomain moire must form.

We further note that the strip dislocations in this device are buried ~15 layers below the surface and yet nucleate at order of magnitude lower loads than in previous experiment that studied surface dislocations (without the superlubricant medium) by applying ~ 40 µN loads [35]. Also, we find no evidence for flexoelectric effects that prefer down-pointing polarization due to the concave nature of the loading force[56].

In some cases of a single dislocation strip between two domains (Fig. 3c), we could accurately push and reposition the strip within the cavity. Here, a uni-directional scanning of the fast axes perpendicular to the strip axes does not change the domain pattern (see straight arrows scan illustration and unaltered potential map). In contrast, the strip precisely follows the end of the scanning pattern position for fast axis with both back and forward scanning along the strip ($3^{rd}$, $4^{th}$ images in Fig. 3c).

**Conclusions and outlook:** The measured switching responses of the polytype islands imply device concepts that were practically out of reach to date. Note that any interlayer shift by one atomic spacing switches the cavity color in the potential map, making the array an exceptionally sensitive strain detector. The polarization is also extremely sensitive to twists and out-of-plane motions, appearing only once the layers perfectly sag (see SI. 6 and Fig. S7). Robust formation of single domain polytypes under finite global twist and electric field switching enable multiferroic devices with coercive fields that depend on the cavity dimensions and the elastic coupling to adjacent cavities in the array. We expect similar responses in other binary compounds hosting additional electronic order, including magnetism, that couples to the structural phase and provides a rich multiferroic response[10]. While the semi-metallic graphene spacer assists with imaging the surface potential in these experiments, switching it to an insulating substance (like twisted *h*-BN) and covering the system with a conducting electrode (like graphene) is straightforward. In this case, each cavity acts as a multiferroic tunneling transistor as thin as two atomic layers. Shaping the cavities into non-circular shapes or tuning the symmetry and dimensions of the array are versatile tools to control multi-switching and collective phase transitions. We expect a similar switching response in polar and non-polar polytypes of graphitic multilayers[27] that exhibit rich correlated states at low temperatures. Here, we envision dense arrays of polytype islands, further interacting by extended quasi-particles that may propagate in the functional layers between the cavities. The ability to design, stabilize, switch, and couple nm-scale islands of distinct periodic crystals should mark a valuable engineering milestone in layered vdW crystals.

## Methods

**Sample preparation:** Our vdW cavity arrays include (top to bottom): *h*-BN / top functional layer / graphene spacer / bottom functional layer / graphite gate electrode / $SiO_2$ substrate assembled using dry polymer stamping method with either monolayers $WSe_2$ or few layers *h*-BN as the functional layers. To etch cavities in the graphene spacers, we used electron-free local anodic oxidation (LAO) lithography (Fig. S2). To form water bubbles between the tip edge and the graphene surface, we maintained a humidity level of ~ 80% using a commercial humidifier inside NX10 AFM from Parksystem. We applied 10 VAC signals to the retracted tip at 40 kHz. After lithography, any remaining oxidized residue was removed by cleaning the surface using HQ:CSC38/Al BS tip B from MikroMasch. The outer spacer dimensions were cut to ~ $10 \times 10$ μm$^2$ using high-intensity pulsed laser light (1064 nm, in a WITEC alpha300 Apyron confocal microscope setup).

**KPFM:** KPFM images were acquired under an inert nitrogen environment using a Park NX10 Hivac system in sideband mode[4]. HQ:NSC35/Pt tip C with ~150 kHz resonant frequency, 5.4 N/m spring constant or N doped diamond tip HQ:DMD-XSC11 tip B with ~110 kHz resonant frequency, 6.5 N/m spring constant were used for KPFM measurements. For sideband measurements, the resonant frequency of the tip was calibrated and then set 2.5 kHz away. The tip was excited by 2 VAC with 9 nm – 15 nm setpoints for all measurements.

**Dislocation control by electric fields:** To maintain a large contact area between the tip and the surface, we used contact scanning with 160 nN on HQ:NSC35/Pt tip C (apex radius below 30 nm), 1 Hz scan speed, and pixel sizes under 5 nm at all specified DC voltage. Alternatively, for HQ:DMD-XSC11 tip B, (apex radius 100-250 nm), we used 90 nN. The same scan direction was used in all scans.

**Dislocation control by tip pressing:** We used HQ:NSC35/Pt tip C with 50 – 600 nN forces, 1 Hz scan speed, and pixel size under 20 nm. The spring constant was calibrated using Sader methods, which is recommended above 100 kHz resonant frequency, and optical lever sensitivity was calibrated using a single-layer graphene on $SiO_2$.

**Aknowledgements**

We thank Neta Ravid and Itzhak Malker for laboratory support. K. W. and T. T. acknowledge support from the JSPS KAKENHI (Grant Numbers 21H05233 and 23H02052) and World Premier International Research Center Initiative (WPI), MEXT, Japan. M.B.S. acknowledges funding by the European Research Council under the European Union's Horizon 2024 research and innovation program ("SlideTronics", consolidator grant agreement no. 101126257) and the Israel Science Foundation under grant nos. 319/22 and 3623/21. We further acknowledge the Centre for Nanoscience and Nanotechnology of Tel Aviv University.

# Switchable Crystaline Islands in Super Lubricant Arrays

## Supplementary Information

Youngki Yeo[1], Yoav Sharaby[1], Nirmal Roy[1], Noam Raab[1], Watanabe Kenji[2], Takashi Taniguchi[3], Moshe Ben Shalom[1]

[1]School of Physics and Astronomy, Tel Aviv University, Tel Aviv, Israel
[2]Research Center for Functional Materials, National Institute for Materials Science, Tsukuba, Japan
[3]International Center for Materials Nanoarchitectonics, National Institute for Materials Science, Tsukuba, Japan

# Contents

# List of Figures

## S1. Device characterization

In the main text we present data from three selected devices.

Device 1, (Figure 1) with a 4.3 nm thick $h$-BN flake used to pick up active 2.3 nm and 2 nm active $h$-BN flakes, and a bilayer graphene spacer. In device 2 (Figure 2) we used 2.5 nm thick $h$-BN to pick up trilayer $h$-BN (1 nm), monolayer graphene spacer, and trilayer $h$-BN. In device 3 (Figure 3) we used a 4.5 nm $h$-BN to pick up monolayer $WSe_2$, graphene spacer and monolayer $WSe_2$. Thicknesses of the Flakes are summarized in Table S1, and an optical microscope image of each device is shown in Figure S1. Pick-up $h$-BN, top layer, graphene spacer, bottom layer are outlined with red, green, blue, and orange lines respectively.

| Device name | $h$-BN thickness (pick-up, top, bottom) | graphene spacer layers |
|---|---|---|
| Device 1: 8.6 nm $h$-BN | 4.3 nm, 2.3 nm, 2.0 nm | Bilayer |
| Device 2: 4.5 nm $h$-BN | 2.5 nm, 1 nm, 1 nm | Monolayer |
| Device 3: $WSe_2$ | 4.5 nm, monolayer $WSe_2$ | Trilayer |

**Table S1. Flakes thickness of three devices presented in the main text.**

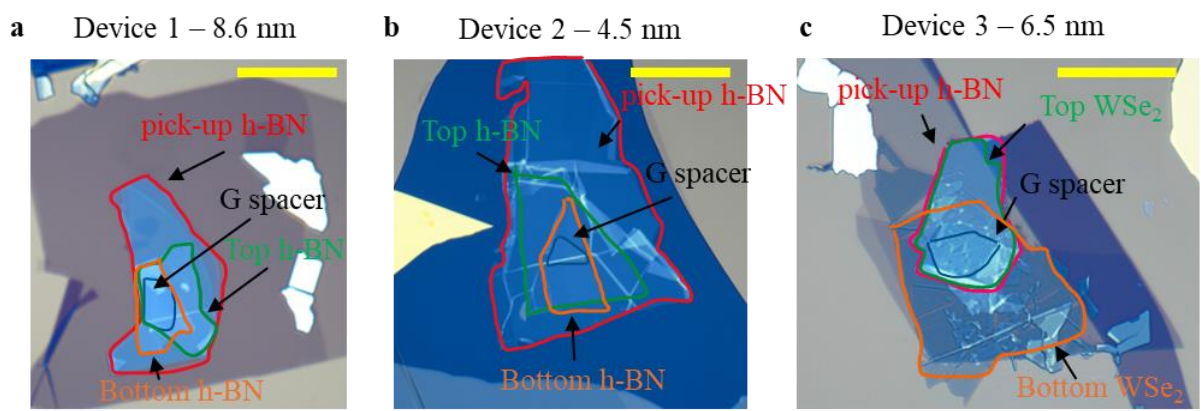

**Figure S1. Optical device images. a,b,c**, Optical microscope image of devices 1,2,3 presented in figures 1,2,3 in the main text respectively. Scale bars indicate 25 μm.

## S2. Graphene spacer lithography

The quality of the cavity edges is crucial to minimize the pinning of dislocation boundary strips. Our best results were obtained using the electrode free local anodic oxidation (LAO) lithography method[1]. Step-by-step procedures are summarized as follows,

1. Choose the wrinkle- and tape-residue free graphene monolayer with optical microscope.

2. Pt-coated AFM tip (Mikromasch HQ:NSC35/Pt Tip C) was used to clean $10\times10$ μm$^2$ lithographic area in advance for removing dirt which is not captured by optical microscopy. Relative humidity (RH) was maintained 60-80 % to boost chemical reaction between tip and graphene with commercial humidifier.

3. Tip was retracted from the graphene surface to form water bubbles on the tip edge. 10 VAC, 40 kHz were applied in PFM mode (NX10, Parksystem *inc.*).

4. Using the lithographic option in NX10, predesigned bitmap image was used to make various shapes. 100 nN setpoint was used for each pixel approach. We avoid higher setpoint results in breaking and flipping graphene edge.

A typical AFM topography map and an optical microscopy image of a trilayer graphene spacer are shown in Figure S2a, b. The cavity diameter is reduced down to ~20 nm which is related to the tip apex dimensions (see AFM map) and remains free of oxidation residues. In contrast, cutting the spacer with high-intensity pulsed laser light resulted in substantial oxidation rings (see bright rings in Fig. S2c) that damaged the device performance. We note that various laser exposure times and beam intensities always resulted in oxidized layers and topographic eruptions at the edges, extending by approximately 1 μm from the center of the heated area. These oxide layers are not visible in optical microscopy (Figure S2d).

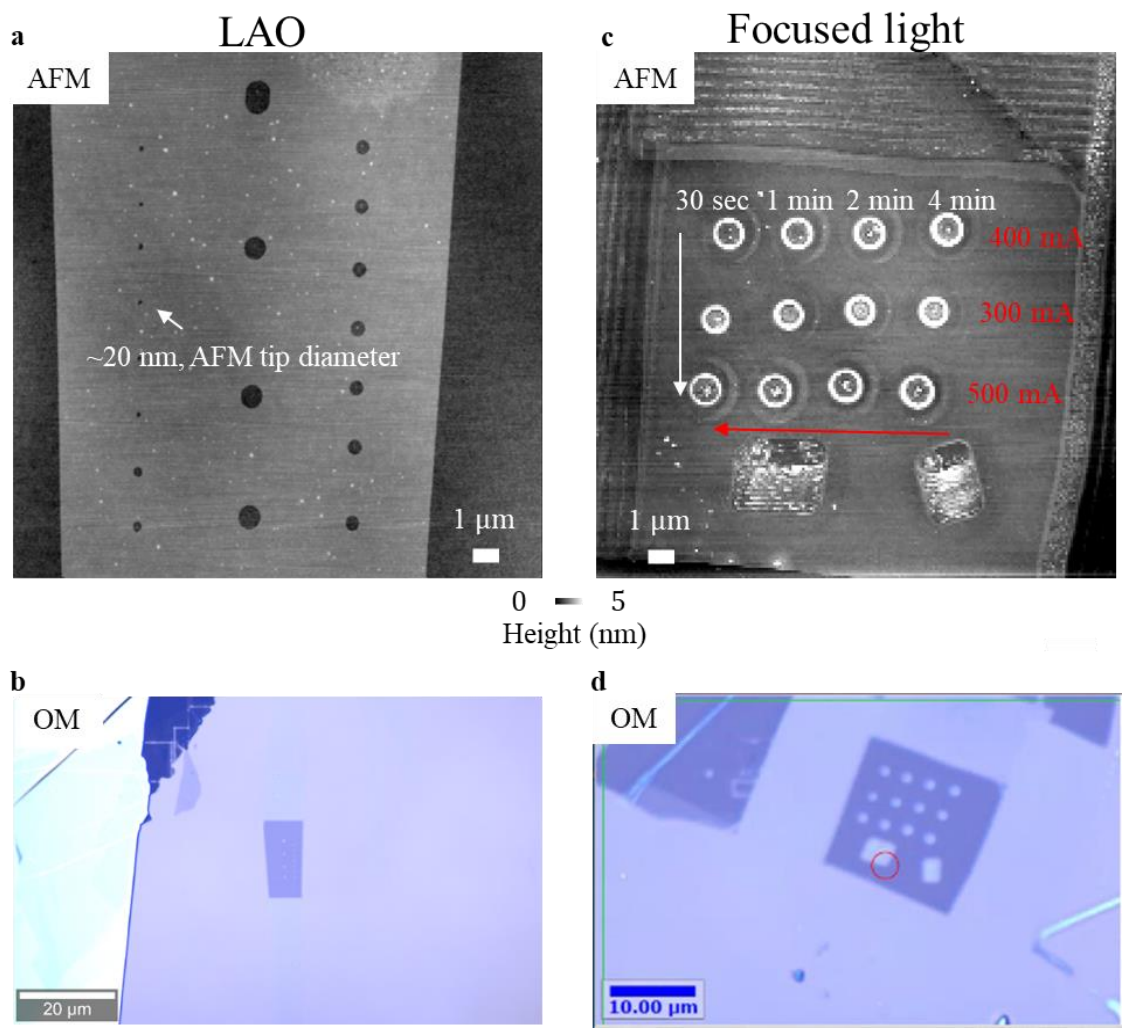

**Figure S2. Comparison of spacer lithography methods. a,b,** AFM and optical microscope image of local anodic oxidation lithography with cavity diameter down to ~20 nm. **c,d,** AFM and optical image of laser pulse lithography.

## S3. Size dependence of coercive switching fields

Switching a uniform single-domain cavity into a full oppositely-polarized state involves intermediate domain patterns that govern the system response. To reveal the intermediate state pattern, we imaged the cavity's surface potential evolution after each increment of the external electric field. Figure S3 shows a selected set of images for a 150 nm cavity (see the hysteresis loop of this cavity in Fig. 2 main text). In this case, we used bias increments of ~0.3 V, corresponding to displacement field steps of 0.06 V/nm, and measured repeatable hysteresis loops for 5 times. The first two maps show a fully-polarized cavity after polling beyond the positive/negative coercive fields (as indicated). Starting with a uniform up (bright) polarization in Fig. 2b in the main text, the following image (1 V, at time $t_3$) shows the first bias in which a boundary wall nucleates and a dark domain covering ~1/2 of the cavity area appears. This domain pattern remains the same up to a 3 V bias, where the cavity switches to a fully down-polarized state. The following maps show reversible switching along hysteresis loops with a coercive field between -2.1 $V_{tip}$ and -2.4 $V_{tip}$ for up-polarization and a coercive field between 2.7 $V_{tip}$ and 3.0 $V_{tip}$ for down-polarization.

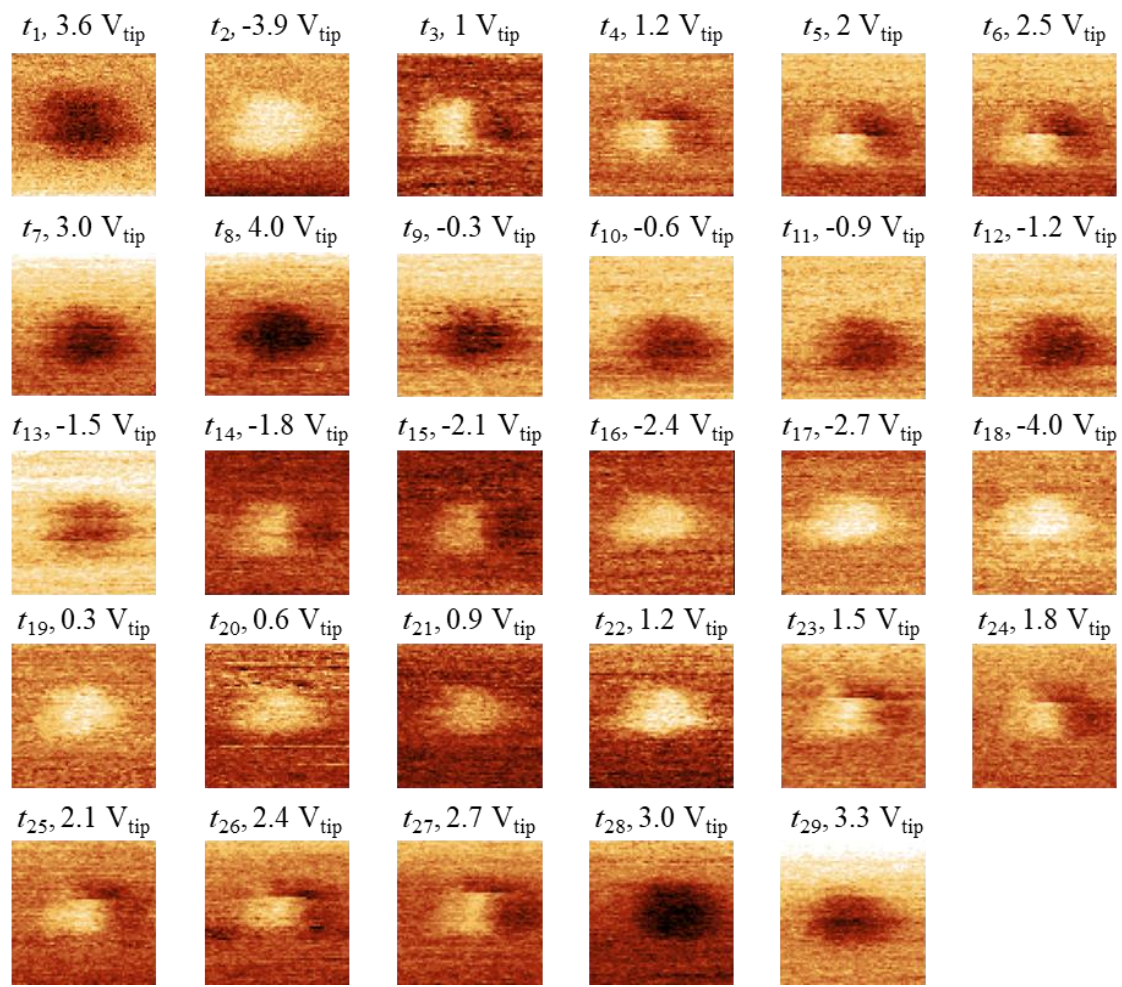

**Figure S3. Imaging ferroelectric hysteresis loop for a 150 nm cavity (Device 2, see Fig.2 main text).** Domain pattern imaged by KPFM after bias scans at the voltage and time indicated on each map.

Figure S4 presents hysteresis loop measurements of the three additional cavities marked by dashed frame in Fig. 2a of the main text. The additional 150 nm-size cavity and the 250 nm-size cavities exhibit a switching bias window of 5 V, while the 350 nm-size cavity does not switch completely in a window of 8 V. We could switch up-polarization at −4 V but could not achieve complete down-polarization even at 4.5 V. We note that higher field values were avoided due to finite damage observed (in other cavities). We also note an asymmetry in positive and negative coercive fields in different cavities, while the overall window remains the same. Since the center of the switching window shifts from positive to negative potential in different cavities, and seems to be smaller for cavities that are further separated from neighbors and pining contaminants, we attribute this anisotropy to elastic coupling of the cavity to its environment, rather than vertical electric fields arising by flexoelectric response[2].

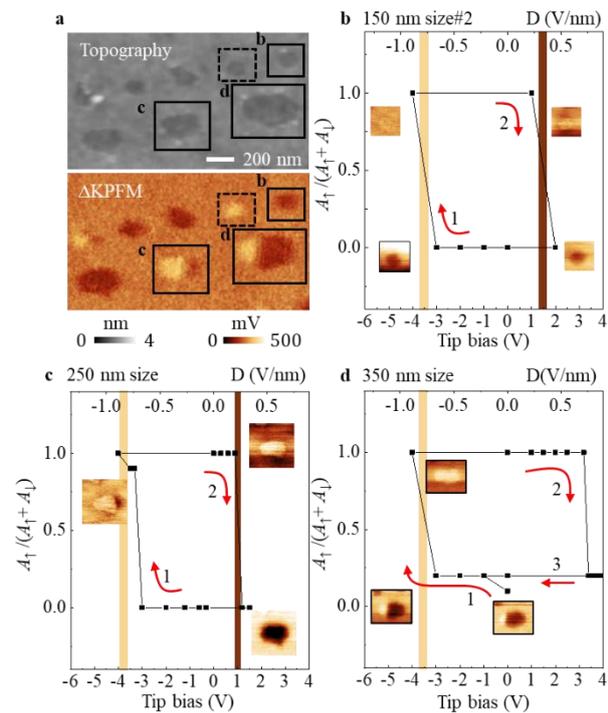

**Figure S4. Additional ferroelectric hysteresis measurements (Device 2). a**, Topographic and KPFM maps of the cavity array. **b**, **c**, **d**, Hysteresis loops of the cavities marked in **a**, with diameters 150, 250, 350 nm respectively. Bright and Dark column stands for the switching to complete up and down polarization.

## S4. Intermediate switching states

Here, a cavity of 250 nm diameter is scanned with a sharp tip (< 30 nm apex diameter) in Figure S5, allowing better spatial resolution and a more local polling electric field. A biased scan of the single domain map (dark uniform potential, panel **b**) first opens a 120° triangle domain (bright, panel **c**), which then extends by 60° to a straight dislocation strip (boundary of dark/bright, panel **d**). We note that the tip force and bias affect the stability of the intermediate state and that so far we were not able to robustly control the triangular intermediate states. We call for further experiments to establish the intermediate domain pattern as a function of the cavity dimensions, shape, and elastics coupling to adjacent cavities.

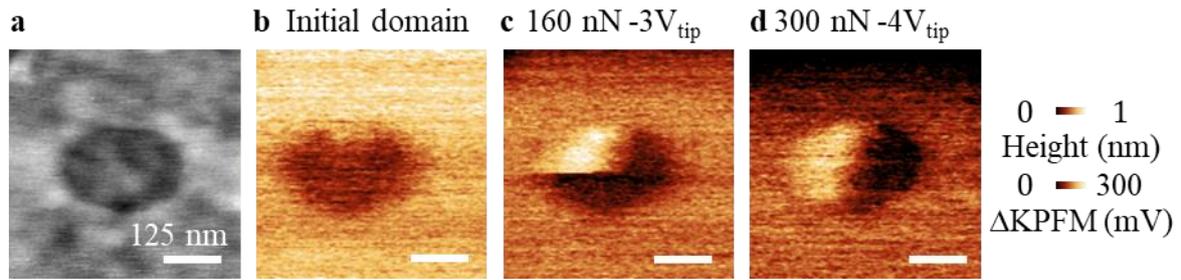

**Figure S5. Triangular intermediate transition states (Device 2). a**, Topographic image of a 250 nm cavity. **b,c,d**, KPFM image measured before and after applying bias scans as indicated in each map.

## S5. Threshold force for boundary strip sliding.

To test the threshold pressure for dislocation nucleation and movement, we conducted dragging experiments with unbiased tip. Figure S6 shows the potential maps of cavity arrays after applying increasingly growing forces to a narrow tip of less than 30 nm apex radius. The first map showing any domain wall motion appears after applying a 200 nN force (marked by blue rectangular frames). Note that variations in the cavity matrix potential are extremely sensitive to any deformation in the array. Contact scanning with 200 nN in the opposite direction reverts the domain pattern in a reversible manner, while higher force levels moved the strips in additional cavities and to larger extent (see 250 nN maps). The latter indicates different threshold values in different cavities as naturally expected. Applying forces below the 200 nN threshold value and after doing dragging experiments, did not change the potential pattern as expected (see bottom two maps).

We used HQ:NSC35/Pt – Tip C for these experiments, with typical spring constant value of 5.4 N/m. Assuming that 20 nm tip apex curvature, threshold pressure is approximately 161 MPa.

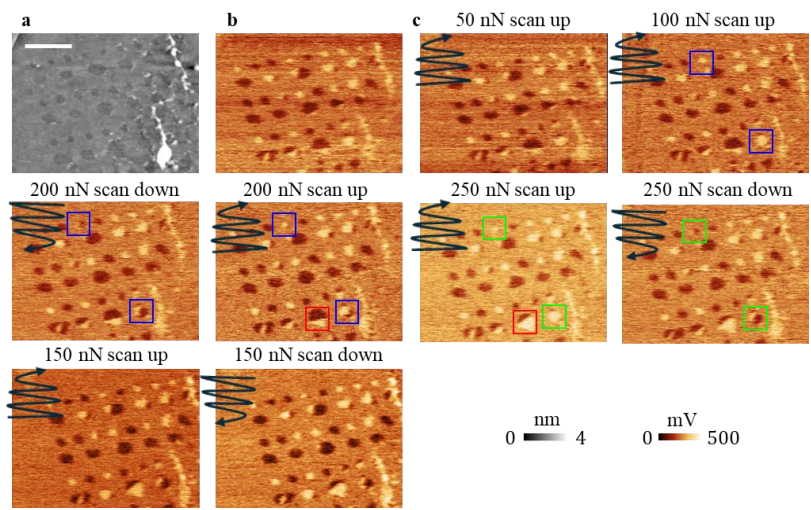

**Figure S6. Determination of threshold pressure for dislocation movement. a**, Topographic AFM image of the measured array. **b**, KPFM signals before tip pressing. **c**, Successive pressing experiments. Scanning illustration and applied force indicate the contact mode scan parameters before the KPFM image.

## S6. Sagging of active layers in pristine cavities

Occasionally, the stamping process resulted in commensurate layers sagging and the appearance of clear polar polytypes only in part of the array. In Figure S7, for example, the topography and surface potential maps show that cavities from the top part of the array are commensurate and polar (see darker cavity color in topography and bright/dark potential domains). Conversely, the bottom part of the array remains non-polar and exhibits flat topography of suspended cavity membranes or pockets of aggregated contaminations (bright topography regions). In this case, we scanned the array in contact mode with a finite pressure of 300 nN, pushing the active layers to dwell into the vdW adhesion and to push away the self-cleaning contamination pockets. Imaging the array after the several pressured scans confirms clean and commensurate cavities everywhere (see Fig. S7c,d). Hence, we note that the surface potential signal is sensitive to the vertical adhesion and the planar interlayer motion.

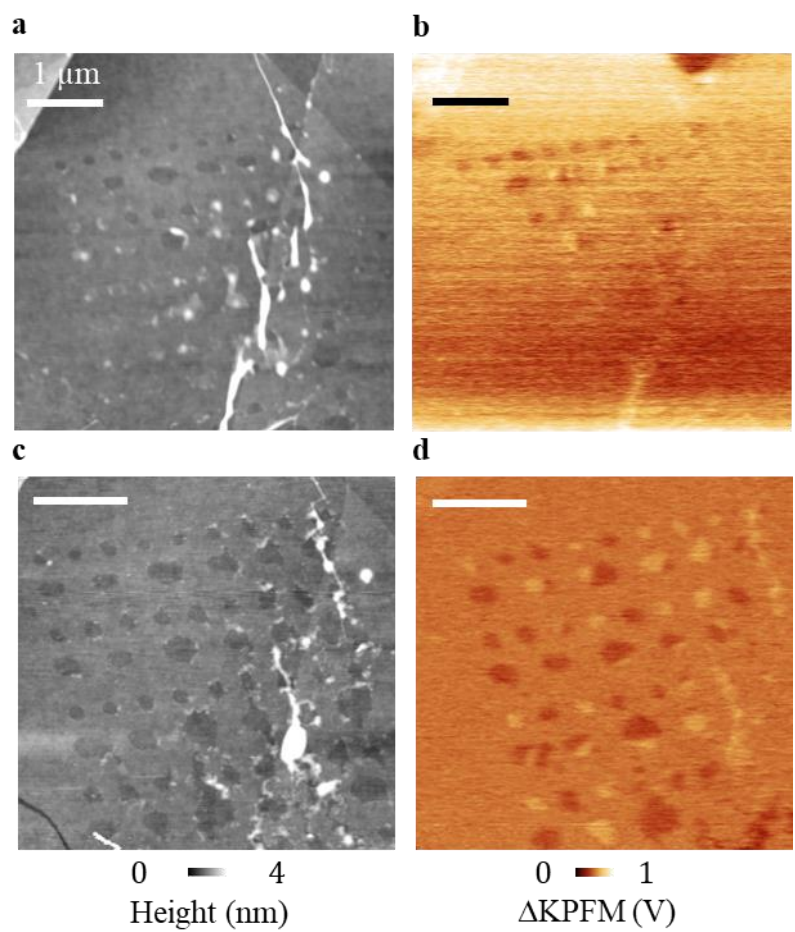

**Figure S7. Imaging sagging of active layers (Device 2). a,b,** Topography and surface potential maps of as-fabricated cavity array. **c,d**, Topographic and KPFM maps after AFM contact mode scans at a pressure of 300 nN. Scale bars are 1 μm.

## S7. Effect of shear/friction force around the cavity

To understand detailed impact of mechanical switching of dislocations inside the cavities, it is essential to compare the shear forces applied at the edges and within the cavity. The friction force induces tortional movement of the tip which is reflected in difference between forward and backward lateral photodiode signals (V). We calibrated conversion factor (nN/V) of lateral force by $SiO_2$ substrate where the friction coefficient is known as 0.08 (ref.[3]). Our estimated $h$-BN friction coefficient is 0.055 derived from the conversion factor which is also reported in AFM friction experiment on $h$-BN[4]

Friction to normal load conversion curve measured on atomically flat $h$-BN flakes is presented in figure S8a. A topographic and friction force maps are shown in panel b,c respectively taken around a cavity at a 300 nN loading force. The brighter signal in c indicates that the lateral friction forces are higher at the edge of cavity, reaching ~30 nN compared ~15 nN in the flat regions. Both values are much below the values reported in previous experiments of mechanical domain nucleation in uniform crystalline flakes (with loading forces exceeding µN (ref.[5])).

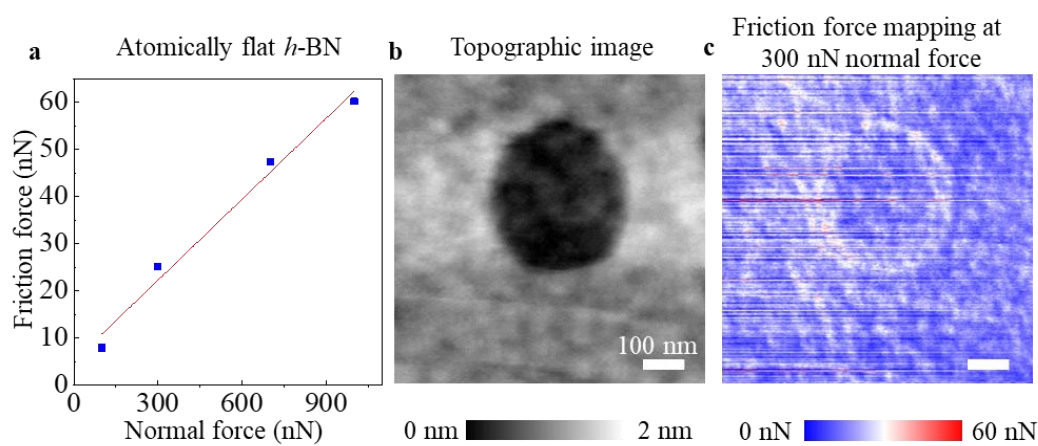

**Figure S8. Friction force measurement around the cavity. a,** Friction force calibration curve measured in atomically flat h-BN. **b**, Topographic image of cavity in Device 3 measured in 300 nN loading force. **c**, Corresponding friction force mapping at 300 nN normal force applied.